\documentclass{emulateapj}
\usepackage{natbib,aas_macros}
\usepackage{amsmath,graphicx,psfrag,float,color,booktabs}
\usepackage{ulem}

\newcommand{\teff}{$T_{\rm eff}$}
\newcommand{\tc}{$T_{\mathrm{C}}$}
\newcommand{\ME}{$M_{\oplus}$}
\newcommand{\MS}{$M_{\odot}$}
\newcommand{\hda}{HD\,20782}
\newcommand{\hdb}{HD\,20781}
\newcommand{\hdab}{HD\,20782/81}

\newcommand{\cyga}{16\,Cyg\,A}
\newcommand{\cygb}{16\,Cyg\,B}

\begin{document}

\submitted{Accepted for publication in The Astrophysical Journal}

\title{Detailed Abundances of Planet-Hosting Wide Binaries.\ I.\
Did Planet Formation Imprint Chemical Signatures in the Atmospheres of \hdab\altaffilmark{*}?}

\altaffiltext{*}{The data presented herein were obtained at the Las Campanas Observatory with
the Magellan/MIKE spectrograph.}

\author{
Claude E. Mack III\altaffilmark{1}, 
Simon C. Schuler\altaffilmark{2}, 
Keivan G. Stassun\altaffilmark{1,3}, 
Joshua Pepper\altaffilmark{4,1}, 
and John Norris\altaffilmark{5} 
}

\affil{
\altaffiltext{1}{Department of Physics and Astronomy, Vanderbilt University, Nashville, TN 37235, USA; claude.e.mack@vanderbilt.edu}
\altaffiltext{2}{University of Tampa, 401 W Kennedy Blvd, Tampa, FL 33606, USA}
\altaffiltext{3}{Department of Physics, Fisk University, Nashville, TN 37208, USA}
\altaffiltext{4}{Department of Physics, Lehigh University, Bethlehem, PA 18015, USA}
\altaffiltext{5}{Research School of Astronomy \& Astrophysics, The Australian National University, Weston ACT 2611, Australia.}
}

\begin{abstract}
Using high-resolution, high signal-to-noise echelle spectra
obtained with Magellan/MIKE,
we present a detailed chemical abundance analysis of both stars 
in the planet-hosting wide binary system \hda\ + \hdb.
{Both stars are G dwarfs, and presumably coeval, forming in the same
molecular cloud. Therefore we expect that they should possess {the same} bulk metallicities.}
{Furthermore, both stars also host  giant planets on eccentric orbits with pericenters $\lesssim$0.2 AU.
Here, we investigate if planets with such orbits could lead to the host stars ingesting material, which in turn
may leave similar chemical imprints in their atmospheric abundances.}
We derived abundances of 15 elements spanning a range of condensation 
temperatures (\tc\ $\approx$ 40--1660~K). 
{The two stars are found to have a mean element-to-element abundance difference
of {$0.04\pm0.07$ dex}, which is consistent with both stars having identical bulk metallicities.}
{In addition,} for both stars, the refractory elements 
(\tc\ $> 900$ K) exhibit a positive correlation 
{between abundance (relative to solar)}
{and} \tc, 
{with similar slopes of $\approx$1$\times$10$^{-4}$ dex K$^{-1}$.}
The measured positive correlations 
{are not perfect; both stars exhibit a scatter of 
$\approx$5$\times$10$^{-5}$ dex K$^{-1}$ about the mean trend, and certain
elements (Na, Al, Sc) are similarly deviant in both stars.}
{These findings}
are discussed in the context of models for giant planet migration that
predict the accretion of H-depleted rocky material by the host star.
{We show that a simple simulation of a solar-type star 
accreting material with Earth-like composition predicts}
{a positive---but} 
{imperfect---correlation between
refractory elemental abundances and \tc.}
{Our measured slopes 
are consistent with what is predicted for the} 
{ingestion} 
{of 10--20 Earths}
{by each star in the system.}
{In addition, the specific element-by-element scatter might
be used to distinguish between planetary accretion and Galactic
chemical evolution scenarios.}
\end{abstract}


\section{INTRODUCTION}\label{s:intro}
Exoplanet surveys like NASA's {\it Kepler} mission are
discovering planets in a variety of environments, e.g., systems with multiple stellar components,
which suggests that planet formation mechanisms are remarkably robust. 
An important result in attempts to understand these planet formation
mechanisms is that giant planets are found to be more prevalent around
solar-type stars that are typically enriched in metals by
$\sim$0.15 dex relative to similar stars that have no detected 
giant planets~\citep[e.g.,][]{2005ApJ...622.1102F,2010ApJ...720.1290G}.
This evidence indicates that giant planet formation is most successful
in metal-rich environments. 

Beyond overall metallicity, investigations of abundance patterns in elements besides
Fe in planet-hosting stars have uncovered evidence that planet hosts may be enriched or depleted
(relative to the Sun) with elements of high condensation temperatures (\tc\ $\gtrsim$ 900 K, i.e.,
the refractory elements that are the major components of rocky planets) 
depending on the architecture and evolution of the their planetary systems. 
There are at least two planet formation processes that may alter stellar surface abundances:
(1) the accretion of hydrogen-depleted rocky material~\citep{1997MNRAS.285..403G}, which would result in the
{\it enrichment} of the stellar atmosphere, and 
(2) H-depleted rocky material in terrestrial planets may be withheld
from the star during their formation, which would result in the {\it depletion} 
of heavy elements relative to H in
the stellar atmosphere~\citep{2009ApJ...704L..66M}. 
For the enrichment scenario,~\citet{2011ApJ...732...55S}
suggest that stars with close-in giant planets ($\sim$0.05 AU) may be more enriched with elements of high
condensation temperature (\tc). 
{This is thought to be a result of giant planets which form in the
outer planetary system migrating inward  to their present close-in positions.}
As they migrate, they can push
rocky material into the host star~\citep[e.g.,][]{2008ApJ...673..487I, 2011A&A...530A..62R}. 
For the depletion scenario, \citet{2009ApJ...704L..66M}
and \citet{2009A&A...508L..17R} propose that the depletion of refractory elements in Sun-like stars
may correlate with the presence of terrestrial planets. 
Certainly there are processes other
than planet formation that may alter stellar atmospheric abundances,
but these effects can be mitigated by simultaneously considering
a pair of stars that have experienced essentially the same evolution and environments over the course of their
lives, {such as stars in wide binaries.}

Indeed, wide stellar binaries known to harbor planets are {valuable} laboratories for studying the 
connection between how planets form and the chemical compositions of their host stars. Since most binary 
stars are believed to have formed coevally from a 
common molecular cloud~\citep[][and references therein]{2011ASPC..447...47K}, 
planet-hosting wide binaries are particularly valuable, because both stars can be presumed to have the 
same age and initial composition.  
{In fact, \citet {2004A&A...420..683D,2006A&A...454..581D} studied the differential Fe abundances
for a set of 50 wide binaries. They found that only one binary pair possessed a $\Delta$[Fe/H] $>$ 0.09~dex, 
while for the majority of the systems they found $\Delta$[Fe/H] $<$ 0.03~dex. 
Thus, for components of wide binaries where at least one star possesses a planet, it is reasonable to expect that
any significant difference in their present-day chemical abundances is most likely due to some
aspect of the planet formation process.}

For example, the investigation by~\citet{2011ApJ...737L..32S}
of 16\,Cyg (a triple system that includes a wide binary pair of two nearly identical stars,
plus the secondary hosts a giant planet at $\sim$1.7 AU while the primary does not) found that \cyga\ and
\cygb\ were chemically identical (However, we should note that \citet{2011ApJ...740...76R}
found that \cyga\ is more metal rich than \cygb\ by {$0.041\pm0.007$ dex}, but \citet{2012ApJ...748L..10M}
found that the two stars are chemically identical). 
The authors speculated that one possible reason \cygb\ formed a giant
planet, while \cyga\ may not have, is because \cyga\ itself has a resolved M dwarf companion (the tertiary in
the system). This third star may have truncated the primary's circumstellar disk and inhibited planet formation
\citep[e.g.,][]{1996ApJ...458..312J,2005MNRAS.363..641M}. Since the two stars must be the same age,
and in addition they were found to be chemically identical, the authors were forced to consider 
{the properties of the system described above, which}
could have led to these two stellar twins failing to form planetary systems with similar architectures. The 16\,Cyg
wide binary was an ideal first system for this kind of comparison study, because the component stars have almost
identical physical properties, i.e., their masses are nearly equal. This minimizes systematic errors 
that may arise from analyzing two stars with drastically different basic stellar properties~\citep{2011ApJ...737L..32S}.

Ascertaining how planet formation may influence the composition of host star atmospheres could
revolutionize target selection for future exoplanet surveys. If chemical abundance patterns can identify
a star as a planet host, then a single high-resolution spectrum$-$instead of solely relying on large,
time-intensive monitoring surveys$-$will permit selection of probable planet hosts among nearby stars in our
Galaxy. Furthermore, if particular chemical signatures indicate the existence of specific kinds of planets, such
as terrestrial planets, considerably more targeted searches for Solar System analogs would be possible.

The goal of this series of papers is to study the interplay between planet formation and
the chemical composition of the host star by directly comparing the chemical abundances of each stellar
pair in planet-hosting wide binaries.
{This paper presents the analysis of detailed abundance trends in the two stars comprising the \hdab\ system.}
{\hda\ and \hdb\ are a common proper motion wide binary with an
angular separation of $252''$ and a projected physical separation of $\sim9,000$ AU {\citep{2007A&A...462..345D,2009A&A...494..373M}}.
They are both solar-type stars with spectral types of G1.5V and
G9.5V, {and apparent V magnitudes of 7.36 and 8.48}, respectively~\citep{2006AJ....132..161G}.}

For \hdab\, we present the investigation of the only known binary 
system where both stars have detected {planets.}
\hda\ has a Jupiter-mass planet on a very eccentric ($e\sim0.97$) orbit at $\sim$1.4 AU~\citep{2006MNRAS.369..249J}, {and}
\hdb\ hosts two moderately eccentric ($e\sim0.1-0.3$) Neptune-mass planets within $\sim$0.3 AU
{(M. Mayor 2013, private communication)}. 
{Therefore, {\it if} the formation and evolution of planetary systems with different
architectures affect the host star composition in distinct ways, studying systems like \hdab\ allows us to discern
which aspects of their architectures play the most important roles.}

In Section~\ref{s:data}, we describe our observations, reductions, and spectral analysis.
In Section~\ref{s:results}, we summarize the main results, 
{including the finding that both stars in \hdab\ exhibit similar
positive trends between refractory elemental abundance and \tc.}
In Section~\ref{s:disc}, we discuss the results
in the context of previous studies and a simple calculation that predicts how the accretion of Earth-like
rocky planets would affect refractory elemental abundances as a function of \tc.
{We find that the observed trends between refractory elemental
abundance and \tc, and the element-by-element scatter relative to the mean
trends, are consistent with the ingestion by both stars of
10--20 Earths}. Finally, in Section~\ref{s:conc}
we highlight our main conclusions.

\section{DATA AND ANALYSIS}\label{s:data}

\begin{figure*}
\centering
\includegraphics[scale=0.9]{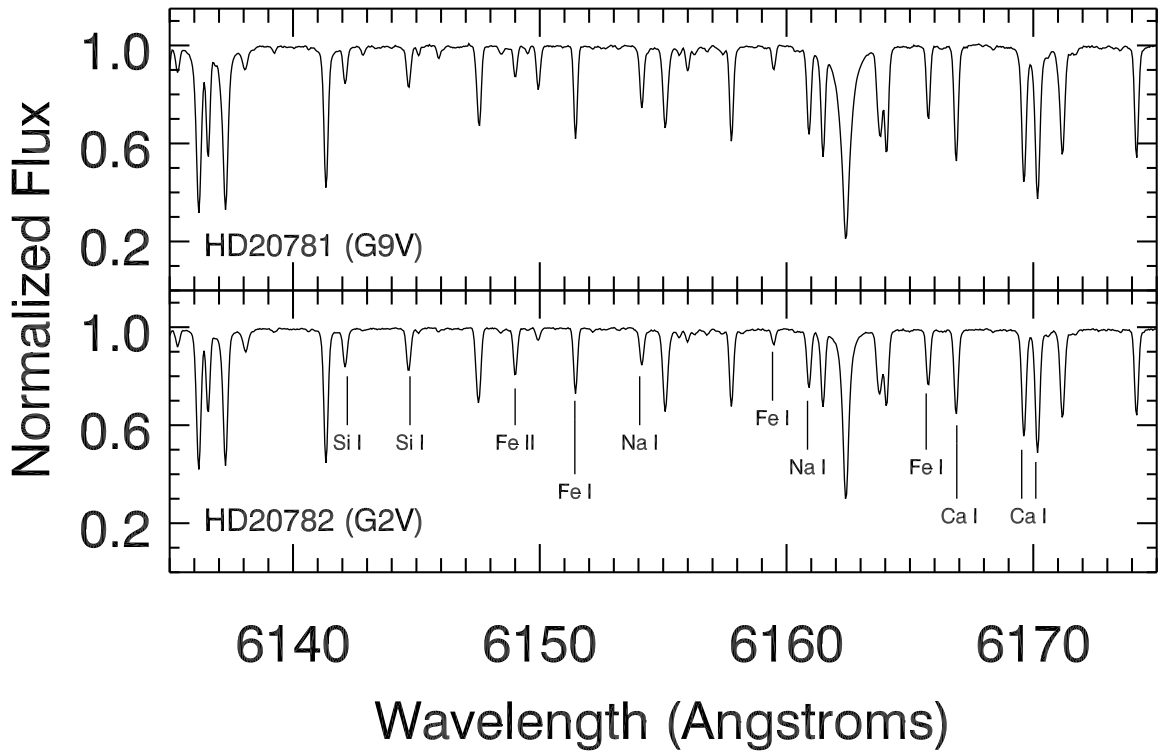}
\caption{Sample Magellan/MIKE spectra for \hdab, spanning the wavelength range from $\lambda6135-\lambda6175$.}
\label{fig:spec}
\end{figure*}

\begin{figure}
\centering
\includegraphics[scale=0.4]{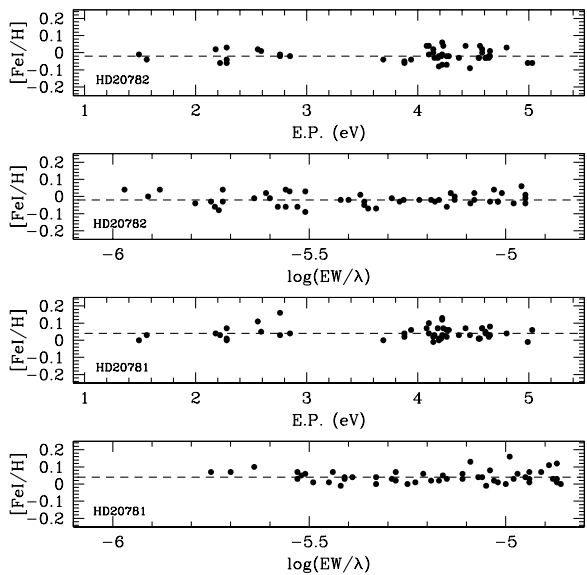}
\caption{Plots of [FeI/H] vs excitation potential and reduced equivalent 
width for both \hdab. The dashed lines indicate the mean values of [FeI/H],
which are $-0.02$ dex and $+0.04$ dex for \hda\ and \hdb, respectively.}
\label{fig:fe1_corr}
\end{figure}

{For both \hdab, on UT 2012 Feb 08 we obtained} high-resolution, 
high-signal-to-noise ratio (SNR) spectra with the 6.5-m 
Magellan II (Clay) telescope~\citep{2003SPIE.4837..910S} and MIKE echelle spectrograph~\citep{2003SPIE.4841.1694B}.
{The spectra covered a wavelength range from $\sim3500-9500$\AA.}
Three exposures were taken of both \hda\ and \hdb, 
with a total integration time of 540s for \hda\ and 1200s for \hdb. 
Multiple bias frames and flat
field exposures were taken at the beginning of the night. A Thorium-Argon lamp exposure
was taken at the beginning and end of the night for wavelength calibration. 
{The data were reduced using standard {\sc iraf} routines.}

{The final reduced spectra possess a resolution of $R=\lambda / \Delta \lambda \sim40,000$
and SNR in the continuum region near $\lambda 6700$ of $\sim$600 for \hdb\ and $\sim$620 for \hda.}
Sample spectra spanning the wavelength region $\lambda$6135-$\lambda$6175 are
shown in Figure~\ref{fig:spec}.
{A solar spectrum (sky) was also obtained for
derivation of relative abundances, and has an SNR of $\sim$610
near $\lambda 6700$.}

\begin{deluxetable}{lcccc}
\tablecolumns{5}
\tablewidth{0pt}
\tablecaption{Stellar Parameters \& Abundances\tablenotemark{a}\label{tab:params}}
\tablehead{
	\colhead{}&
	\colhead{}&
	\colhead{\hda}&
	\colhead{}&
	\colhead{\hdb}
	}
\startdata
$T_{\mathrm{eff}}$ (K) && $5789 \pm 38$          && $5324 \pm 52$   \\
$\log g$ (cgs)         && $4.41 \pm 0.12$        && $4.51 \pm 0.10$ \\
$\xi$ (km s$^{-1}$)    && $1.32 \pm 0.10$        && $1.02 \pm 0.11$ \\
$[$C/H$]$   \dotfill   && $-0.07 \pm 0.03\tablenotemark{b} \pm 0.04$\tablenotemark{c} && $-0.09 \pm0.03 \, \pm 0.04$ \\
$[$O/H$]$   \dotfill   && $+0.04 \pm 0.01 \, \pm 0.05$ && $-0.08 \pm0.04 \, \pm 0.07$ \\
$[$Na/H$]$  \dotfill   && $-0.08 \pm 0.02 \, \pm 0.03$ && $-0.06 \pm0.03 \, \pm 0.05$ \\
$[$Mg/H$]$  \dotfill   && $+0.04 \pm 0.01 \, \pm 0.06$ && $+0.10 \pm0.00 \, \pm 0.05$ \\
$[$Al/H$]$  \dotfill   && $-0.01 \pm 0.00 \, \pm 0.02$ && $+0.04 \pm0.00 \, \pm 0.03$ \\
$[$Si/H$]$  \dotfill   && $-0.02 \pm 0.01 \, \pm 0.02$ && $-0.01 \pm0.01 \, \pm 0.02$ \\
$[$Ca/H$]$  \dotfill   && $+0.04 \pm 0.01 \, \pm 0.05$ && $+0.06 \pm0.03 \, \pm 0.06$ \\
$[$Sc/H$]$  \dotfill   && $-0.03 \pm 0.02 \, \pm 0.06$ && $+0.04 \pm0.02 \, \pm 0.05$ \\
$[$Ti/H$]$  \dotfill   && $+0.06 \pm 0.01 \, \pm 0.05$ && $+0.12 \pm0.02 \, \pm 0.06$ \\
$[$V/H$]$   \dotfill   && $-0.01 \pm 0.01 \, \pm 0.04$ && $+0.10 \pm0.02 \, \pm 0.07$ \\
$[$Cr/H$]$  \dotfill   && $-0.05 \pm 0.01 \, \pm 0.03$ && $+0.03 \pm0.02 \, \pm 0.05$ \\
$[$Mn/H$]$  \dotfill   && $-0.01 \pm 0.06 \, \pm 0.07$ && $+0.04 \pm0.05 \, \pm 0.07$ \\
$[$Fe/H$]$  \dotfill   && $-0.02 \pm 0.01 \, \pm 0.02$ && $+0.04 \pm0.01 \, \pm 0.03$ \\
$[$Co/H$]$  \dotfill   && $-0.05 \pm 0.01 \, \pm 0.03$ && $+0.03 \pm0.01 \, \pm 0.04$ \\
$[$Ni/H$]$  \dotfill   && $-0.02 \pm 0.01 \, \pm 0.03$ && $+0.01 \pm0.01 \, \pm 0.03$ \\

\enddata

\tablenotetext{a}{Adopted solar parameters: \teff\ $=5777$ K, $\log g=4.44$, 
and $\xi=1.38$ km s$^{-1}$.}
\tablenotetext{b}{{$\sigma_{\mu}=\sigma / \sqrt{N-1},$ where $\sigma$ is the standard deviation and $N$ is the number
of lines measured for a given element.}}
\tablenotetext{c}{$\sigma_{Total}$-- quadratic sum of $\sigma_{\mu}$ and uncertainties in the elemental abundance resulting from uncertainties in \teff, $\log g$, and $\xi$.}

\end{deluxetable}

\begin{deluxetable*}{lcccccccrcccrcccrcc}
\tablecolumns{16}
\tablewidth{\textwidth}
\tablecaption{Lines Measured, Equivalent Widths, and Abundances\label{tab:linelist}}
\tablehead{
     \colhead{}&
     \colhead{}&
     \colhead{$\lambda$}&
     \colhead{}&
     \colhead{$\chi$}&
     \colhead{}&
     \colhead{}&
     \colhead{}&
     \colhead{}&
     \colhead{}&
     \colhead{}&
     \colhead{}&
     \multicolumn{3}{c}{\hda}&
     \colhead{}&
     \multicolumn{3}{c}{\hdb}\\
     \cline{13-15} \cline{17-19}\\
     \colhead{Element}&
     \colhead{}&
     \colhead{(\AA)}&
     \colhead{}&
     \colhead{(eV)}&
     \colhead{}&
     \colhead{$\log \mathrm{gf}$}&
     \colhead{}&
     \colhead{EW$_{\odot}$}&
     \colhead{$\log N_{\odot}$}&
     \colhead{$\log N_{\odot,\mathrm{synth}}$\tablenotemark{a}}&
     \colhead{}&
     \colhead{EW}&
     \colhead{$\log N$}&
     \colhead{$\log N_{\mathrm{synth}}$}&
     \colhead{}&
     \colhead{EW}&
     \colhead{$\log N$}&
     \colhead{$\log N_{\mathrm{synth}}$}
     }
\startdata
\ion{C}{1} && 5052.17 && 7.68 && -1.304 &&  36.2 & 8.51 & \nodata &&  36.1 & 8.50 & \nodata &&  17.2 & 8.35 & \nodata \\ 
\ion{C2}{0} && 5086 && \nodata && \nodata &&  \nodata & \nodata & 8.43 &&  \nodata & \nodata & 8.32 &&  \nodata & \nodata & 8.35 \\ 
\ion{C2}{0} && 5135 && \nodata && \nodata &&  \nodata & \nodata & 8.44 &&  \nodata & \nodata & 8.36 &&  \nodata & \nodata & 8.43 \\ 
\ion{C}{1} && 5380.34 && 7.68 && -1.615 &&  22.8 & 8.54 & \nodata &&  20.8 & 8.47 & \nodata &&  10.9 & 8.44 & \nodata \\ 
\ion{O}{1} && 6300.30 && 0.00 && -9.717 &&   5.5 & 8.69 & \nodata &&   6.3 & 8.72 & \nodata &&   6.5 & 8.52 & \nodata \\ 
\ion{O}{1} && 7771.94 && 9.15 &&  0.369 &&  65.6 & 8.77 & \nodata &&  71.2 & 8.84 & \nodata &&  34.7 & 8.71 & \nodata \\ 
\ion{O}{1} && 7774.17 && 9.15 &&  0.223 &&  57.8 & 8.79 & \nodata &&  60.5 & 8.82 & \nodata &&  30.4 & 8.74 & \nodata \\ 
\ion{O}{1} && 7775.39 && 9.15 &&  0.001 &&  46.5 & 8.80 & \nodata &&  47.5 & 8.81 & \nodata &&  23.1 & 8.75 & \nodata \\ 
\ion{Na}{1} && 5682.63 && 2.10 && -0.700 && 119.9 & 6.52 & \nodata && 107.9 & 6.41 & \nodata && 135.1 & 6.41 & \nodata \\ 
\ion{Na}{1} && 6154.23 && 2.10 && -1.560 &&  38.2 & 6.31 & \nodata &&  33.2 & 6.23 & \nodata &&  49.2 & 6.23 & \nodata \\ 
\ion{Na}{1} && 6160.75 && 2.10 && -1.260 &&  58.1 & 6.31 & \nodata &&  53.9 & 6.26 & \nodata &&  76.4 & 6.31 & \nodata \\ 
\ion{Mg}{1} && 5711.09 && 4.35 && -1.833 && 100.8 & 7.56 & \nodata && 102.0 & 7.59 & \nodata && 128.3 & 7.66 & \nodata \\ 
\ion{Mg}{1} && 6841.19 && 5.75 && -1.610 &&  64.1 & 7.85 & \nodata &&  66.0 & 7.89 & \nodata &&  74.6 & 8.14\tablenotemark{b} & \nodata \\ 
\ion{Al}{1} && 6696.02 && 3.14 && -1.347 &&  36.7 & 6.24 & \nodata &&  34.9 & 6.23 & \nodata &&  52.7 & 6.28 & \nodata \\ 
\ion{Al}{1} && 6698.67 && 3.14 && -1.647 &&  20.7 & 6.21 & \nodata &&  19.8 & 6.20 & \nodata &&  32.6 & 6.25 & \nodata \\ 
\enddata
\tablenotetext{a}{Indicates the $\log{N}$ abundance determined from the synthetic fit to a given line.}
\tablenotetext{b}{{The $\log{N}$ abundance for this line was rejected as spurious, as described in paragraph 4 of Section~\ref{s:data}.}}
\tablecomments{This table is published in its entirety in the online journal.  
A portion is shown here for guidance regarding 
its form and content.}

\end{deluxetable*}     

In each star, abundances of 15 elements have been derived from
{the observed spectra.}
The 2010 version of the LTE spectral analysis package {\sc moog} \citep{1973ApJ...184..839S}
was used to perform the spectral analysis. The 
abundances were derived from measurements of the equivalent widths (EWs) of atomic lines
using the {\sc spectre} analysis package~\citep{1987BAAS...19.1129F}. 
{We adopted our line list from~\citet{2011ApJ...732...55S}.}
Stellar parameters 
were obtained by requiring excitation and ionization balance of 
the \ion{Fe}{1} and \ion{Fe}{2} lines {in the standard way.}
{Plots of [FeI/H] vs. excitation potential and reduced equivalent
width are provided in Figure~\ref{fig:fe1_corr}, which shows that the
correlations are zero as required.}
{The atomic excitation energies ($\chi$) and transition probabilities ($\log{\mathrm{gf}}$)
were taken from the Vienna Atomic Line Database~\citep[VALD;][]{1995A&AS..112..525P,1999A&AS..138..119K}}.
{For each element, the abundances were determined relative to solar via a line-by-line
differential analysis.}

Carbon abundances are also derived 
with the {\it synth} driver in {\sc moog} to synthetically fit the C$_2$ features
at $\lambda5086$ and $\lambda5135$.  Oxygen abundances were determined with the {\sc moog} {\it blends}
driver for the forbidden line at $\lambda6300$, and EW measurements of 
the near-infrared triplet
at $\lambda7771$, $\lambda7774$, and $\lambda7775$.
{Also, we suspect that
the \ion{Mg}{1} line at $\lambda6841.19$ is blended with a line that becomes stronger
in stars with \teff$\lesssim5400$~K, and thus we rejected the Mg abundance it
yielded for \hdb\ as spurious.}

For the odd-Z elements V, Mn, and Co, the abundances of which can
be overestimated due to hyper-fine structure (hfs) effects {\citep{2000ApJ...537L..57P}},
spectral synthesis incorporating hfs components has been used to verify
the EW-based results.  {The hfs components for these elements were obtained
from \citet{2006ApJ...640..801J}, and the line lists for wavelength regions encompassing each
feature were taken from VALD.} The adopted V, Mn, and Co abundances are derived from
the hfs analysis and those lines with EWs that were not significantly altered
by hfs.  

The abundance and error analyses {for all elements} are described in detail in
\citet{2011ApJ...732...55S}.  
The stellar abundances (relative to the solar abundances 
derived from the solar spectrum), parameters, and uncertainties 
for \hdab\ are summarized in Table \ref{tab:params}.  The adopted line list, 
EWs, and line-by-line abundances of each element for  
\hdab\ and the Sun are given in Table \ref{tab:linelist}.

\section{RESULTS}\label{s:results}
\begin{figure}
\centering
\includegraphics[scale=0.4]{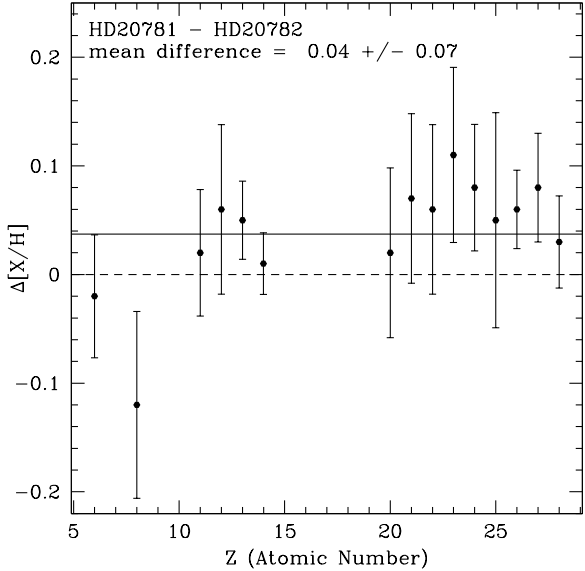}
\caption{Differential abundances for \hdab\ as a function of atomic number (Z).
The solid line represents the mean difference of $0.04\pm0.07$ dex, and the dashed line is
meant to guide the eye at 0.0 dex.}
\label{fig:diff}
\end{figure}

As shown in Table \ref{tab:params}, the stellar parameters we determined for \hdab\ 
are consistent with the primary being a $\sim$G2V and the secondary being a $\sim$G9.5V. 
The differences in parameters (in the sense of primary minus secondary) 
are {: $\Delta$\teff\ $=+465 \pm
64$ K, $\Delta \log g=-0.10 \pm 0.16$ dex, and $\Delta \xi=+0.30\pm 0.15$ 
km s$^{-1}$}.
{Furthermore, according to the PASTEL catalogue of stellar parameters~\citep{2010A&A...515A.111S},
the mean literature values for the stellar parameters of \hda\ (\teff$\sim5800$~K, $\log{g}\sim4.4$~dex,
and [Fe/H] $\sim-0.06$~dex) are in good agreement with ours. 
For \hdb, our values agree with the mean literature values for \teff\ ($\sim$5300~K) and $\log{g}$~($\sim$4.4~dex),
but there is a considerable spread of $\sim$0.2 dex ($-0.18$---$+0.01$~dex) in the published [Fe/H] values for this star.
The upper end of this range is consistent with the value of [Fe/H] that we derive for \hdb.}
The abundances of the 15 individual elements are shown graphically in 
Figure \ref{fig:diff}. The abundance differences shown in Figure \ref{fig:diff} 
are the means of the line-by-line differences for each element.  
{The mean abundance difference is $0.04\pm0.07$ dex,}
as expected for coeval stars in a binary system.

\begin{figure}
\centering
\includegraphics[scale=0.4]{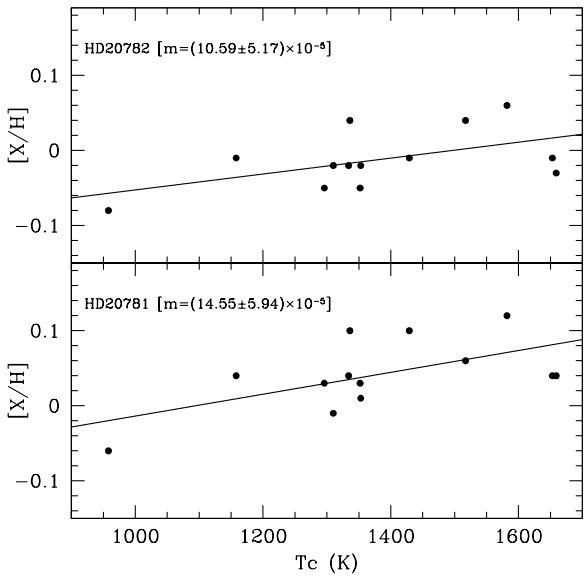}
\caption{{\it Unweighted} linear fits to abundance vs.  condensation temperature (\tc) for \hdab. 
}
\label{fig:tc_unweight}
\end{figure}

The abundances of \hdab\ are shown versus \tc\ in Figures~\ref{fig:tc_unweight}~and~\ref{fig:tc_weight}.  
The condensation temperatures were taken from the 50\% \tc\ values
derived by \citet{2003ApJ...591.1220L}. 
Only the refractory elements (\tc$\gtrsim 900$ K) are displayed, because it is among 
these elements that the chemical signature of planet formation {has been shown} to be strongest 
\citep{2009ApJ...704L..66M}.  We performed both unweighted and weighted linear fits
{to the [X/H] versus \tc\ abundance relations to investigate possible correlations.}
{For our analysis and discussion, we adopt the {\it weighted} fits. 
However, in Figure~\ref{fig:tc_unweight}, we provide the unweighted fits 
for comparisons to previous studies that only reported unweighted fits.}

{As can be seen in Figures~\ref{fig:tc_unweight}-\ref{fig:tc_weight}, 
{the} slopes of the unweighted linear least-squares fits {are}:}
$m_{\mathrm{82}}= (10.59 \pm 5.17) \times 10^{-5}$ dex K$^{-1}$ and 
$m_{\mathrm{81}}= (14.55 \pm 5.94) \times 10^{-5}$ dex K$^{-1}$ for \hda\ and
\hdb\, respectively.  {The} slopes of the weighted linear least-squares fits {are}:
$m_{\mathrm{82}}= ( 9.71 \pm 4.57) \times 10^{-5}$ dex K$^{-1}$ and 
$m_{\mathrm{81}}= (13.60 \pm 6.57) \times 10^{-5}$ dex K$^{-1}$ for \hda\ and
\hdb\, respectively.
{Thus the correlation between refractory elemental abundance
and \tc\ is not perfect, with individual elements exhibiting scatter 
relative to the mean trend.}
{Nonetheless,}
the slopes of the weighted linear fits to refractory abundances vs \tc\ are 
{modestly} statistically significant ($\sim$2$\sigma$).
In addition, both a Pearson's $r$ and a Kendall's $\tau$ correlation test 
indicate that the abundances and \tc\ are correlated at
$>$ 90\% confidence for both stars (Pearson $r$ confidence of 97\% for
\hdb\ and 92\% for \hda).

In the discussion that follows,
we consider this result in the context of a model to predict the degree to which we might
expect a modest correlation between abundance and \tc\ from host stars that have ingested a small
amount of rocky planetary material.

\vspace{3em}
\section{DISCUSSION}\label{s:disc}

\subsection{How well correlated are abundance\\ and \tc\ expected to be?}
\label{sec:sim_acc}

To estimate the impact that the accretion of a rocky planet would have on the atmospheric composition of a
solar-type star, we simulated the accretion of a massive body with Earth-like composition
onto the Sun.
{Since Jupiter and Saturn are predominantly composed of H and He with approximately
solar H/He ratios~\citep{2003NewAR..47....1Y,2007prpl.conf..591L}, the chemical composition of
a gas giant planet is likely to be fairly similar to the protoplanetary disk,
and thereby to the host star. The accretion of gas giant planets, therefore,
would be unlikely to produce the refractory element versus \tc\ correlation observed
in some planet hosts. Such trends would be expected to arise only from the
accretion of H-depleted rocky material.}

{To perform our calculation of the expected trend between refractory
elemental abundance and \tc,}
we begin by considering 
what would happen to the refractory abundances of the Sun if it accreted
{a certain multiple of} \ME\ of refractory material with a composition similar to the Earth. 
{Since the metallicities of \hdab\ are consistent with solar 
(all elemental abundances are within $\sim$0.1 dex of their solar values)
it is reasonable to assume that the primordial
abundances of both stars were similar to the present-day solar abundances.}
We use the values of
\citet{McDonough01} to obtain the mass fraction for each element in the Earth. 
With the mass of the
Earth, and the molar mass for each element, we can determine an absolute number of atoms for each
element. Then we add this amount of each element into the solar convection zone and see how the
abundances change.
Given the mass of the Sun, the mass fraction of hydrogen, and the fact that at 30 Myr 
{(by which time gas should have dissipated from
the protoplanetary disk, and only fully formed planets and a debris disk remain)} 
the convection zone was 3\% of the Sun's mass~\citep{1993ApJ...418..457S}, 
we can determine the amount of hydrogen in the Sun's convection
zone at that time. Using the solar abundances listed in~\citet{2009ARA&A..47..481A}, 
{the photospheric abundance of each element
relative to hydrogen can be determined. Thus, the change in the abundance of each element due to the accretion of 
{a certain multiple of} \ME\ of Earth-like rocky material can be calculated}.

\begin{figure}
\centering
\includegraphics[scale=0.45]{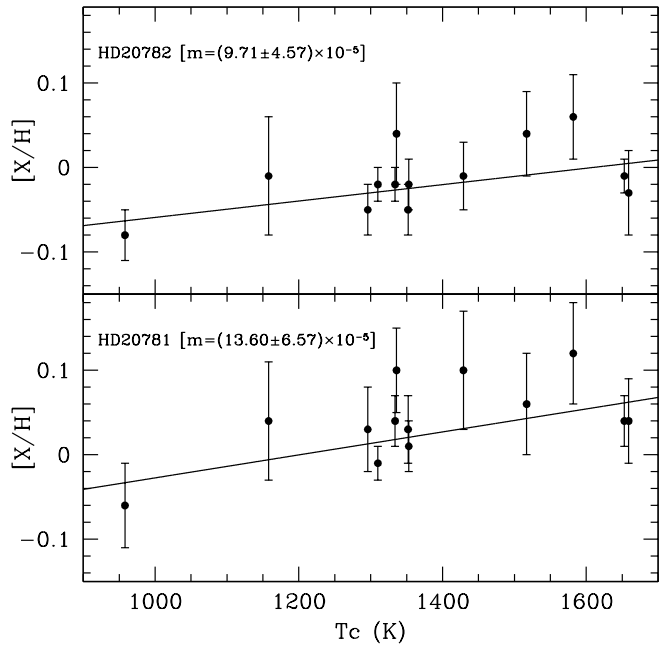}
\caption{{\it Weighted} linear fits to abundances vs. condensation temperature (\tc) for \hdab. 
}
\label{fig:tc_weight} 
\end{figure}

\begin{figure}
\centering
\includegraphics[scale=0.48]{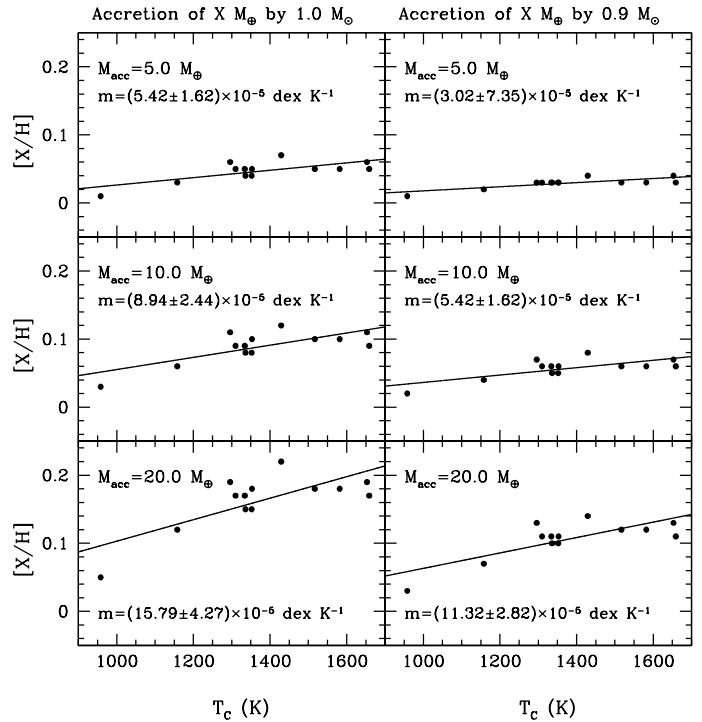}
\caption{{\it Unweighted} linear fits to {simulated} abundances vs.\ condensation temperature (\tc) from our {modeled} 
accretion of $X$ amount of \ME\ by a solar-composition star (see Section~\ref{sec:sim_acc}). {The {\it left} panel shows
the results for the accretion of 5, 10, and 20 \ME\ by a 1.0 \MS\ star, and the {\it right} panel shows the results for
the accretion of the same three amounts of Earth-like material, but for a 0.9 \MS\ star.}}
\label{fig:tc_acc} 
\end{figure}

{Through the order-of-magnitude calculation described above,
we derived the values for the [X/H]-\tc\
slopes in each of the accretion scenarios shown in Figure~\ref{fig:tc_acc}.}
For example, if a solar-type star were to accrete 5\ME of material with a chemical composition
similar to the Earth, then 
{from our calculation} we would expect a trend with \tc\ among the refractories that corresponds to
a slope of $(5.42\pm1.62)\times10^5$ dex K$^{-1}$ (Figure~\ref{fig:tc_acc}). 
{Furthermore, as a result of our simulation of a Sun-like star accreting Earth-like planets, we would {\it not} expect
a perfect correlation between abundance and \tc. There is some scatter 
about the linear fit to the
simulated data, as shown in Figure~\ref{fig:tc_acc}.  Indeed, some of the elements
deviate in a similar manner from the linear fit in both the observed and modeled data.
{For example,} in Figures~\ref{fig:tc_unweight}$-$\ref{fig:tc_acc}, the elements
Na (\tc\ $\sim$960 K), Al (\tc\ $\sim$1653 K), and Sc (\tc\ $\sim$1659 K)
are consistently below the fits to both the simulated and observed data. 
Since these elements are similarly scattered about the fit in both the model and the observations,
the scatter in the observed correlation may not solely be the result of observational noise.}

{There are several ways to extend the above calculation, for example, to take into account
differences in the mass of the star, the mass-dependence of the size of the convection zone, and 
possible variations in the composition of the planets
accreted by the star. Such additional considerations could perhaps explain additional scatter
in the observed abundances. Here our intent is to illustrate the sense and magnitude of the effect.}
{However, given that $\Delta$\teff\ $=+465 \pm 64$ K for \hdab, we investigated how differences in
the depths of their convection zones would affect the results of our simple model.
Using Figure~1 from Pinsonneault et al.\ (2001), which provides a relationship between \teff\ and the mass of the
convection zone, we estimate that for a star like \hdb\ (\teff$\sim5300$ K) the mass of the convection
zone is at most $\sim$0.05\MS. Using the empirical relationships described in Torres et al.\ (2010)---which 
yield the mass and radius of a star as functions of the spectroscopic stellar parameters---we can
estimate the mass of \hdb\ to be $\sim$0.9\MS. This means that about 6\% of the mass of \hdb\ is in the
convection zone (as opposed to $\sim$3\% for stars with masses similar to the Sun). Thus, while our simulation shows that the
ingestion of 10\ME\ of Earth-like material would produce the measured slope for \hda, about twice
as much material is required to produce the measured slope for \hdb (Figure~\ref{fig:tc_acc}).}

\subsection{Interpretation of the positive slopes\\ for \hdab\ }
The positive trends with \tc\ seen among the refractory elemental
abundances of \hdab\ {may be} due to the presence of {eccentric} giant planets
that have migrated to orbits within $\sim$1 AU.
\hda\ hosts a very eccentric Jupiter
at 1.4 AU, with a pericenter of $a(1-e)=1.4(1-0.97)\sim0.04$ AU. 
\hdb\ possesses two close-in Neptunes at 0.2 and 0.3~AU.
These giant planets could have pushed refractory-rich planetary material
into their host stars as they migrated inward {to their current orbits}.
{We have shown in Section~\ref{sec:sim_acc} that compared to
a simple model of a solar-type star accreting Earth-like material,
both \hdab\ have slopes that are consistent with the ingestion of $10-20$\ME\ of rocky material.} 

Several studies have performed simulations of giant planet migration that result
in a substantial amount of hydrogen-depleted material falling into 
the star~\citep[e.g.,][]{2008ApJ...673..487I, 2011A&A...530A..62R}. This
usually occurs because of planet-planet scattering in a rocky debris disk after the
gas-rich protoplanetary disk has dissipated.  \citet{2011A&A...530A..62R} found that in 40\% of 
their simulations, giants migrating in debris disk as as result of planet-planet scattering 
removed all rocky material from the planetary system, most of which was accreted by the host star.
Furthermore, for giant planets with a minimum orbital distance less than $\sim$1 AU, all terrestrial
material was destroyed. Both of the planets around \hdb\ have semi-major axes $\lesssim$0.3~AU,
and the eccentric planet hosted by \hda\ has a perihelion distance of only $\sim$0.04~AU.  Therefore the 
simulations performed by~\citet{2011A&A...530A..62R} indicate that
both planetary systems should be devoid of rocky material.

In addition, \citet{2013Natur.493..381K} also noted that over billions of years planets can be driven
into their host stars because of the presence of a wide binary companion and Kozai
resonances. As the binary pair orbits the galaxy, galactic tides can perturb the binary
system and change the pericenter. At closest approach, each star can disrupt any planetary
system that may exist around its binary companion. When~\citet{2013Natur.493..381K} compared
\hdab\ to wide binaries in their simulations with similar masses and semimajor axes, {they
found that $\sim$55\% of the systems like \hdab\ triggered instabilities, and more
that 90\% of those instabilities occurred in the 
planetary system after 100 Myr.} These instabilities resulted in planets colliding with the
star 14\% of the time. Therefore, both perturbations of the stellar binary as well as planetary
migration through planet-planet scattering can lead to the ingestion of planetary material
by the host star, and thereby generate the abundance patterns which are present in
our data and predicted by our accretion model.

\subsection{Comparison to previous work}
Our findings, namely that the positive trends with \tc\ for the refractory elements 
indicate that both \hdab\ have accreted
H-depleted, refractory-rich material, are consistent with the interpretations of 
{\citet{2011ApJ...732...55S, 2011ApJ...737L..32S,2009ApJ...704L..66M,2009A&A...508L..17R}.}
In particular, \citet{2011ApJ...732...55S} analyzed abundances versus \tc\ trends for
10 stars known to host giant planets.  The 
trends with \tc\ for these 10 stars were compared to 
a sample of 121 stars with and without detected giant planets 
from \citet{2010MNRAS.407..314G}; the distribution of slopes with respect to 
[Fe/H] for {the $\sim$120 stars from the \citet{2010MNRAS.407..314G} sample was interpreted} 
as the general trend from Galactic chemical evolution. Of the 10 stars
investigated by \citet{2011ApJ...732...55S}, the
four with very close-in ($\sim$0.05 AU) giant planets
were found to have positive slopes that lie above 
the Galactic trend.  \hdab\ also have positive slopes that lie above this Galactic trend.
The four stars from \citet{2011ApJ...732...55S} were also hypothesized 
to have ingested refractory-rich planetary material as a result of the evolution of 
their planetary systems.
{However, \citet{2013A&A...552A...6G} analyzed a sample of 61 late-F to early-G stars,
29 of which have detected planets and 32 do not. After correcting their
trends with \tc\ for Galactic chemical evolution, they found that their stars with
and without detected planets possessed similar [X/H]-\tc\ slopes. Therefore, they
concluded that, in general, trends with \tc\ may not indicate the presence or absence of planets.}

Part of the value of comparing coeval stars in wide binaries, 
is that any {\it difference} in their abundance trends is most certainly not the result of Galactic chemical processes. 
However, since \hdab\ exhibit {\it similar} abundance trends, this finding could in fact either be the result
of the planet formation process or Galactic chemical evolution.  
{In order to distinguish between these two scenarios, we compared the [X/H]-\tc\
slopes for \hdab\ to the distribution of slopes observed for the
the $\sim$120 stars in the \citet{2010MNRAS.407..314G} sample.
Among the stars within 0.1~dex of solar metallicity in this sample, the slopes of \hdab\ lie in the
upper envelope of the distribution of slopes, which suggests that they are on
the higher end of the Galactic trend.  
Furthermore, the fact that they both deviate from the Galactic trend in the same way
also suggests that their abundances have most likely been changed by a similar process, i.e., 
the accretion of rocky planetary material. 
Finally, it is not obvious that Galactic chemical
evolution can produce the specific element-by-element scatter that we observe 
(e.g., Na, Al, Sc, see Section~\ref{sec:sim_acc}), whereas the planet accretion scenario
appears to reproduce it naturally, at least in our current simple model (Section~\ref{sec:sim_acc}).}

\citet{2009ApJ...704L..66M} and \citet{2009A&A...508L..17R} have performed studies of solar twins (stars with physical parameters nearly
identical to the Sun), and found that solar refractory
abundances decrease as a function of \tc. Therefore, since the Sun formed terrestrial planets, 
they posit that a negative slope may indicate the presence of terrestrial planets, which
contain the refractory-rich, H-depleted material that would have otherwise been accreted by the
host star. Since we can rule out negative slopes for both \hdab\ at the 2$\sigma$ level, the interpretation
suggested by \citet{2009ApJ...704L..66M} and \citet{2009A&A...508L..17R} implies that neither star hosts terrestrial planets, which, as noted previously
for planetary systems with giant planets at $\lesssim$1 AU, is
consistent with models of planet migration that predict that
both \hdab\ are unlikely to host rocky planets~\citep{2005ApJ...620L.111V, 2011A&A...530A..62R}.

Unlike the work performed by~\citet{2011ApJ...732...55S}
on planet-hosting field stars, or the work by \citet{2009ApJ...704L..66M} and \citet{2009A&A...508L..17R} on solar twins culled from a sample of
field stars, or even the work by~\citet{2011ApJ...737L..32S} and \citet{2011ApJ...740...76R} on 16\,Cyg, this paper's focus on \hdab\
permits the comparison of two coeval stars that {\it both} have detected planetary systems. The Kozai
mechanism that is likely the source of the large eccentricity of \cygb\ b~\citep{1997Natur.386..254H,1997ApJ...477L.103M,2005ApJ...627.1001T}, 
is most likely the cause of
the very high eccentricity of \hda\ b as well.  However, in the multiplanetary system hosted
by the secondary star, the two planets \hdb\ b/c can
dynamically interact with each other to
suppress the effect of the Kozai mechanism,
{and prevent highly eccentric orbits~\citep{1997AJ....113.1915I,2011A&A...533A...7B,2011ApJ...742L..24K}.}
{Thus, while the architectures of the planetary systems hosted by \hdab\ are 
{not identical}, the fact that both stars 
possess giant planets with pericenters $\lesssim$0.2 AU probably resulted in the injection of
$10-20$~\ME of Earth-like rocky material into both stars.}

\section{Conclusion}\label{s:conc}

We have performed a detailed chemical abundance analysis of the planet-hosting
wide binary \hdab, which is presently the only known wide binary where both
stars have detected planets. {The mean element-to-element abundance difference
between the two stars is $0.04\pm0.07$ dex, signifying that their bulk metallicities
are identical, as expected for a binary system.} Both stars show 
{modestly} significant ($\sim$2$\sigma$) positive trends with \tc\ 
among their refractory elemental abundances.
{We cannot definitively rule out that these trends may be the result of 
Galactic chemical evolution. However,}
{given the orbital characteristics of the stellar binary,
and the fact that both stars have eccentric giant planets that approach within $\lesssim$0.2 AU, 
models of dynamical interactions between binary stellar companions and models of giant planet migration}
indicate that the host stars could have accreted rocky planetary bodies that would
have initially formed interior to the giant planets. This is
also consistent with previous studies that found positive trends with \tc\ in field
stars with close-in giant planets.

{According to our simple model for the accretion of Earth-like planets,
the slopes of the weighted fits to these trends
are consistent with \hda\ accreting $\sim$10~\ME\
and \hdb\ accreting $\sim$20~\ME\ of material with Earth-like composition.
Our model also predicts that there should not be a perfect correlation between refractory
abundances and \tc\ {for stars accreting H-depleted, rocky planetary material}.}
{Three elements (Na, Al, and Sc) are similarly discrepant with both the fit to the simulated
data and the fit to the observed data.}
Therefore, the scatter in the [X/H]-\tc\ correlation is not necessarily
due solely to {observational noise},
{but may in fact be a signature of the accretion of 
refractory-rich material driven by the inward migration of the 
giant planets orbiting these stars}. 
{Indeed, the specific character of the element-by-element scatter
might be used as a strong discriminant between the planetary accretion and
Galactic chemical evolution scenarios.}
As we investigate other planet-hosting wide binaries, 
we hope to further {refine these} insights into abundances trends 
{and their relation to the planet formation process}.

\acknowledgements
C.E.M. and K.G.S acknowledge support by 
{NSF AAG AST-1009810 and NSF PAARE AST-0849736}.
We also would like to acknowledge Leslie Hebb
for her important contributions to the initial
design and development of the project. Finally,
we thank the anonymous referee for their comments
and suggestions, which significantly improved the
quality of the publication.\\[5pt]

{\it Facilities:} \facility{Magellan:Clay (MIKE spectrograph)}

\newpage


\begin{thebibliography}{50}
\expandafter\ifx\csname natexlab\endcsname\relax\def\natexlab#1{#1}\fi
%
\bibitem[{{Asplund} {et~al.}(2009){Asplund}, {Grevesse}, {Sauval}, \&
  {Scott}}]{2009ARA&A..47..481A}
{Asplund}, M., {Grevesse}, N., {Sauval}, A.~J., \& {Scott}, P. 2009, \araa, 47,
  481
%
\bibitem[Batygin et 
al.(2011)]{2011A&A...533A...7B} Batygin, K., Morbidelli, A., \& Tsiganis, K.\ 2011, \aap, 533, A7 
%
\bibitem[Bernstein et al.(2003)]{2003SPIE.4841.1694B} Bernstein, R., 
Shectman, S.~A., Gunnels, S.~M., Mochnacki, S., 
\& Athey, A.~E.\ 2003, \procspie, 4841, 1694 
%
\bibitem[Desidera et 
al.(2004)]{2004A&A...420..683D} Desidera, S., Gratton, R.~G., Scuderi, S., et al.\ 2004, \aap, 420, 683
%
\bibitem[Desidera et 
al.(2006)]{2006A&A...454..581D} Desidera, S., Gratton, R.~G., Lucatello, S., \& Claudi, R.~U.\ 2006, \aap, 454, 581 
%
\bibitem[Desidera 
\& Barbieri(2007)]{2007A&A...462..345D} Desidera, S., \& Barbieri, M.\ 2007, \aap, 462, 345 
%
\bibitem[{{Fischer} \& {Valenti}(2005)}]{2005ApJ...622.1102F}
{Fischer}, D.~A., \& {Valenti}, J. 2005, \apj, 622, 1102
%
\bibitem[{{Fitzpatrick} \& {Sneden}(1987)}]{1987BAAS...19.1129F}
{Fitzpatrick}, M.~J., \& {Sneden}, C. 1987, in Bulletin of the American
  Astronomical Society, Vol.~19, Bulletin of the American Astronomical Society,
  1129--+
%
\bibitem[Ghezzi et al.(2010)]{2010ApJ...720.1290G} Ghezzi, L., Cunha, K., 
Smith, V.~V., et al.\ 2010, \apj, 720, 1290 
%
\bibitem[{{Gonzalez}(1997)}]{1997MNRAS.285..403G}
{Gonzalez}, G. 1997, \mnras, 285, 403
%
\bibitem[{{Gonzalez} {et~al.}(2010){Gonzalez}, {Carlson}, \&
  {Tobin}}]{2010MNRAS.407..314G}
{Gonzalez}, G., {Carlson}, M.~K., \& {Tobin}, R.~W. 2010, \mnras, 407, 314
%
\bibitem[Gray et al.(2006)]{2006AJ....132..161G} Gray, R.~O., Corbally, 
C.~J., Garrison, R.~F., et al.\ 2006, \aj, 132, 161 
%
\bibitem[Gonz{\'a}lez Hern{\'a}ndez et al.(2013)]{2013A&A...552A...6G} 
Gonz{\'a}lez Hern{\'a}ndez, J.~I., Delgado-Mena, E., Sousa, S.~G., et al.\ 2013, \aap, 552, A6 
%
\bibitem[{{Holman} {et~al.}(1997){Holman}, {Touma}, \& {Tremaine}}]{1997Natur.386..254H}
{Holman}, M., {Touma}, J., \& {Tremaine}, S. 1997, \nat, 386, 254
%
\bibitem[Ida \& Lin(2008)]{2008ApJ...673..487I} Ida, S., \& Lin, D.~N.~C.\ 2008, \apj, 673, 487 
%
\bibitem[Innanen et al.(1997)]{1997AJ....113.1915I} Innanen, K.~A., Zheng, 
J.~Q., Mikkola, S., \& Valtonen, M.~J.\ 1997, \aj, 113, 1915 
%
\bibitem[{{Jensen} {et~al.}(1996){Jensen}, {Mathieu}, \& {Fuller}}]{1996ApJ...458..312J}
{Jensen}, E.~L.~N., {Mathieu}, R.~D., \& {Fuller}, G.~A. 1996, \apj, 458, 312
%
\bibitem[Johnson et al.(2006)]{2006ApJ...640..801J} Johnson, J.~A., Ivans, 
I.~I., \& Stetson, P.~B.\ 2006, \apj, 640, 801 
%
\bibitem[Jones et al.(2006)]{2006MNRAS.369..249J} Jones, H.~R.~A., Butler, 
R.~P., Tinney, C.~G., et al.\ 2006, \mnras, 369, 249 
%
\bibitem[Kaib et al.(2011)]{2011ApJ...742L..24K} Kaib, N.~A., Raymond, 
S.~N., \& Duncan, M.~J.\ 2011, \apjl, 742, L24 
%
\bibitem[Kaib et al.(2013)]{2013Natur.493..381K} Kaib, N.~A., Raymond, 
S.~N., \& Duncan, M.\ 2013, \nat, 493, 381 
%
\bibitem[Kratter(2011)]{2011ASPC..447...47K} Kratter, K.~M.\ 2011, 
Evolution of Compact Binaries, ASPC, 447, 47 
%
\bibitem[Kupka et 
al.(1999)]{1999A&AS..138..119K} Kupka, F., Piskunov, N., Ryabchikova, T.~A., Stempels, H.~C., \& Weiss, W.~W.\ 1999, \aaps, 138, 119 
%
\bibitem[Lissauer 
\& Stevenson(2007)]{2007prpl.conf..591L} Lissauer, J.~J., \& Stevenson, D.~J.\ 2007, Protostars and Planets V, 591 
%
\bibitem[{{Lodders}(2003)}]{2003ApJ...591.1220L}
{Lodders}, K. 2003, \apj, 591, 1220
%
\bibitem[{{Mayer} {et~al.}(2005){Mayer}, {Wadsley}, {Quinn}, \& {Stadel}}]{2005MNRAS.363..641M}
{Mayer}, L., {Wadsley}, J., {Quinn}, T., \& {Stadel}, J. 2005, \mnras, 363, 641
%
\bibitem[McDonough(2001)]{McDonough01} {McDonough}, W. 2001, in The Composition of the Earth,
in Earthquake Thermodynamics and Phase Transitions in the Earth's Interior (International Geophysics
Series, Vol. 76), ed. R. Teisseyre \& E. Majewski (San Diego, CA: Academic Press)
%
\bibitem[{{Mazeh} {et~al.}(1997){Mazeh}, {Krymolowski}, \& {Rosenfeld}}]{1997ApJ...477L.103M}
{Mazeh}, T., {Krymolowski}, Y., \& {Rosenfeld}, G. 1997, \apjl, 477, L103+
%
\bibitem[{{Mel{\'e}ndez} {et~al.}(2009){Mel{\'e}ndez}, {Asplund}, {Gustafsson},
  \& {Yong}}]{2009ApJ...704L..66M}
{Mel{\'e}ndez}, J., {Asplund}, M., {Gustafsson}, B., \& {Yong}, D. 2009, \apjl,
  704, L66
  %
 \bibitem[Metcalfe et al.(2012)]{2012ApJ...748L..10M} Metcalfe, T.~S., 
Chaplin, W.~J., Appourchaux, T., et al.\ 2012, \apjl, 748, L10 
%
\bibitem[Mugrauer 
\& Neuh{\"a}user(2009)]{2009A&A...494..373M} Mugrauer, M., \& Neuh{\"a}user, R.\ 2009, \aap, 494, 373 
%
\bibitem[Pinsonneault et al.(2001)]{2001ApJ...556L..59P} Pinsonneault, 
M.~H., DePoy, D.~L., \& Coffee, M.\ 2001, \apjl, 556, L59 
%
\bibitem[Piskunov et 
al.(1995)]{1995A&AS..112..525P} Piskunov, N.~E., Kupka, F., Ryabchikova, T.~A., Weiss, W.~W., \& Jeffery, C.~S.\ 1995, \aaps, 112, 525 

\bibitem[Prochaska 
\& McWilliam(2000)]{2000ApJ...537L..57P} 
Prochaska, J.~X., \& McWilliam, A.\ 2000, \apjl, 537, L57 
%
\bibitem[{{Ram{\'{\i}}rez} {et~al.}(2009){Ram{\'{\i}}rez}, {Mel{\'e}ndez}, \&
 {Asplund}}]{2009A&A...508L..17R}
{Ram{\'{\i}}rez}, I., {Mel{\'e}ndez}, J., \& {Asplund}, M. 2009, \aap, 508, L17
%
\bibitem[Ram{\'{\i}}rez et al.(2011)]{2011ApJ...740...76R} Ram{\'{\i}}rez, 
I., Mel{\'e}ndez, J., Cornejo, D., Roederer, I.~U., 
\& Fish, J.~R.\ 2011, \apj, 740, 76 
%
\bibitem[{{Raymond} {et~al.}(2011){Raymond}, {Armitage}, {Moro-Mart{\'{\i}}n},
  {Booth}, {Wyatt}, {Armstrong}, {Mandell}, {Selsis}, \&
  {West}}]{2011A&A...530A..62R}
{Raymond}, S.~N., {et~al.} 2011, \aap, 530, A62+
%
\bibitem[Sackmann et al.(1993)]{1993ApJ...418..457S} Sackmann, I.-J., 
Boothroyd, A.~I., \& Kraemer, K.~E.\ 1993, \apj, 418, 457 
%
\bibitem[Shectman \& Johns(2003)]{2003SPIE.4837..910S} Shectman, S.~A., \& Johns, M.\ 2003, \procspie, 4837, 910
%
\bibitem[{{Schuler} {et~al.}(2011a){Schuler}, {Flateau}, {Cunha}, {King},
  {Ghezzi}, \& {Smith}}]{2011ApJ...732...55S}
{Schuler}, S.~C., {Flateau}, D., {Cunha}, K., {King}, J.~R., {Ghezzi}, L., \&
  {Smith}, V.~V. 2011, \apj, 732, 55
  %
\bibitem[Schuler et al.(2011b)]{2011ApJ...737L..32S} Schuler, S.~C., Cunha, 
K., Smith, V.~V., et al.\ 2011, \apjl, 737, L32 
%
\bibitem[{{Sneden}(1973)}]{1973ApJ...184..839S}
{Sneden}, C. 1973, \apj, 184, 839
%
\bibitem[Soubiran et al.(2010)]{2010A&A...515A.111S}
Soubiran, C., Le Campion, J.-F., Cayrel de Strobel, G., \& Caillo, A.\ 2010, \aap, 515, A111 
%
\bibitem[{{Takeda} \& {Rasio}(2005)}]{2005ApJ...627.1001T}
{Takeda}, G., \& {Rasio}, F.~A. 2005, \apj, 627, 1001
%
\bibitem[Torres et 
al.(2010)]{2010A&ARv..18...67T} Torres, G., Andersen, J., \& Gim{\'e}nez, A.\ 2010, \aapr, 18, 67 
%
\bibitem[{{Veras} \& {Armitage}(2005)}]{2005ApJ...620L.111V}
{Veras}, D., \& {Armitage
}, P.~J. 2005, \apjl, 620, L111
%
\bibitem[Young(2003)]{2003NewAR..47....1Y} Young, R.~E.\ 2003, \nar, 47, 1 
\end{thebibliography}
\end{document}